\documentclass[sigconf,natbib=true,anonymous=false]{acmart}
\usepackage{sigir25-llm-sr-demo-frame}

\begin{document}
	\raggedbottom
	\title{AiReview: An Open Platform for Accelerating Systematic Reviews with LLMs}

\author{Xinyu Mao}
\affiliation{%
	\institution{The University of Queensland}
	\city{Brisbane}
	\country{Australia}
}
\email{xinyu.mao@uq.edu.au}

\author{Teerapong Leelanupab}
\affiliation{%
	\institution{The University of Queensland}
	\city{Brisbane}
	\country{Australia}
}
\email{t.leelanupab@uq.edu.au}

\author{Martin Potthast}
\affiliation{%
	\institution{University of Kassel and hessian.AI}
	\city{Kassel}
	\country{Germany}
}
\email{martin.potthast@uni-kassel.de}

\author{Harrisen Scells}
\affiliation{%
	\institution{University of Kassel and hessian.AI}
	\city{Kassel}
	\country{Germany}
}
\email{harry.scell@uni-kassel.de}

\author{Guido Zuccon}
\affiliation{%
	\institution{The University of Queensland}
	\city{Brisbane}
	\country{Australia}
}
\email{g.zuccon@uq.edu.au}

\begin{abstract}
\sloppy
Systematic reviews are fundamental to evidence-based medicine. Creating one is time-consuming and labour-intensive, mainly due to the need to screen, or assess, many studies for inclusion in the review. Several tools have been developed to streamline this process, mostly relying on traditional machine learning methods. Large language models (LLMs) have shown potential in further accelerating the screening process. However, no tool currently allows end users to directly leverage LLMs for screening or facilitates systematic and transparent usage of LLM-assisted screening methods. This paper introduces (i) an extensible framework for applying LLMs to systematic review tasks, particularly title and abstract screening, and (ii) a web-based interface for LLM-assisted screening. Together, these elements form AiReview---a novel platform for LLM-assisted systematic review creation. AiReview is the first of its kind to bridge the gap between cutting-edge LLM-assisted screening methods and those that create medical systematic reviews. The tool is available at \url{https://aireview.ielab.io}. The source code is also open sourced at \url{https://github.com/ielab/ai-review}.

\end{abstract}

\maketitle
\setcounter{footnote}{0} 

	\begin{figure}[t!]
	\centering
	\includegraphics[width=\columnwidth]{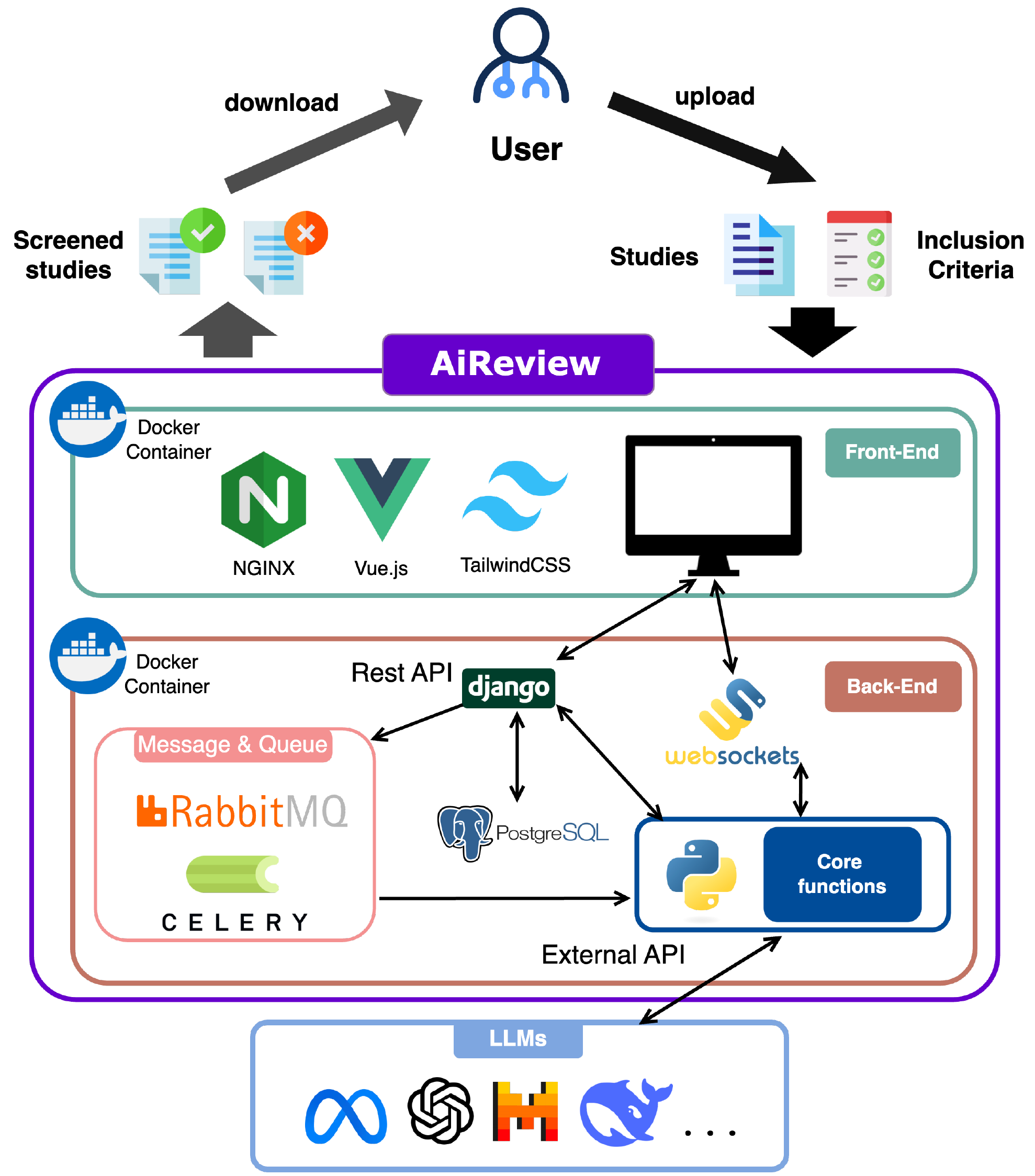}
	\caption{Holistic architecture and workflow of AiReview.}
	\label{fig:arch}
\end{figure}

\begin{figure*}[t!]
	\centering
	\begin{subfigure}[b]{0.641\textwidth}  
		\includegraphics[width=\textwidth]{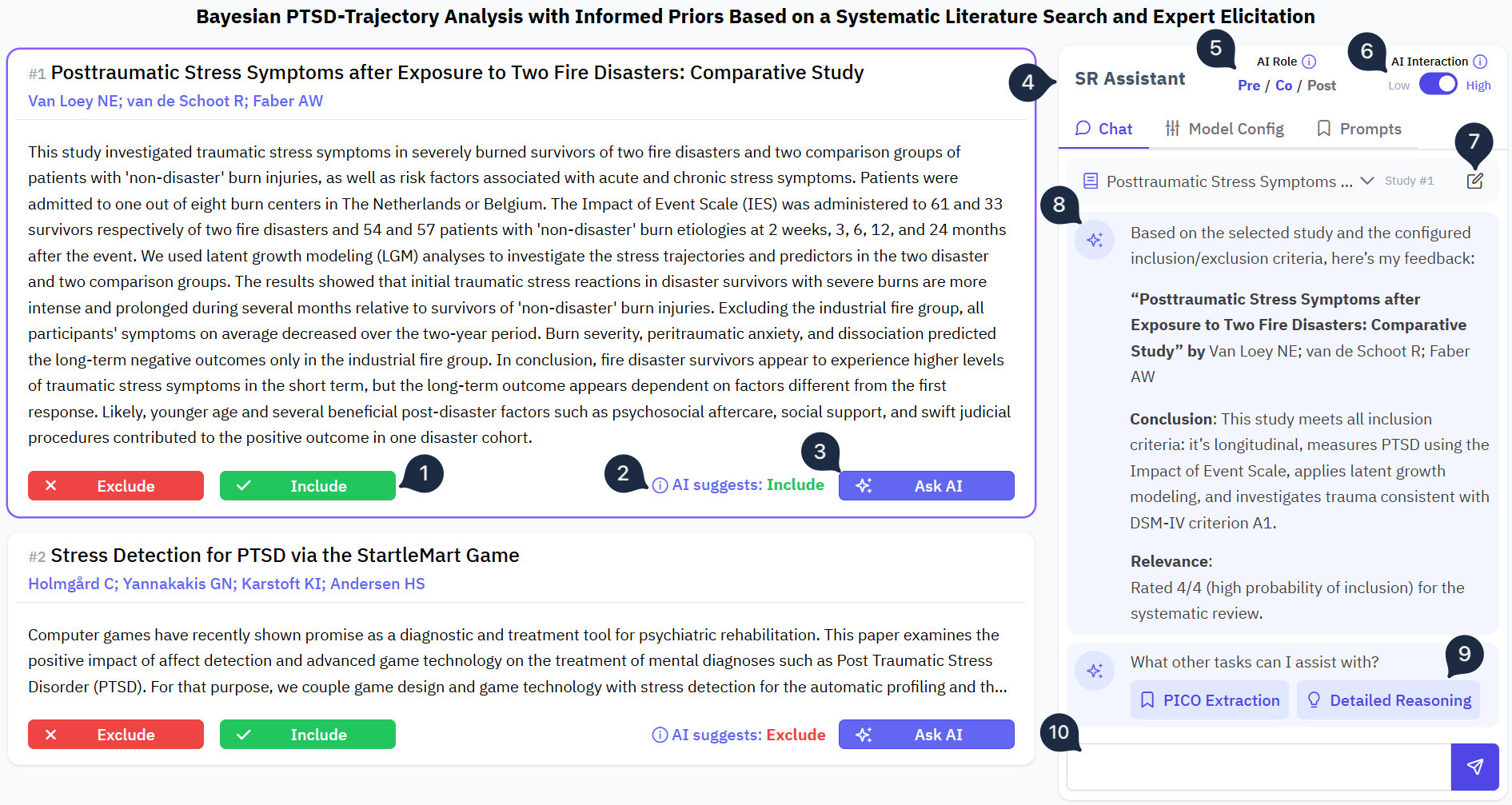}
		\caption{AiReview screening interface with the SR Assistant Panel}
		\label{fig:main}
	\end{subfigure}
	\hspace{1pt}
	\begin{subfigure}[b]{0.1715\textwidth}  
		\includegraphics[width=\textwidth]{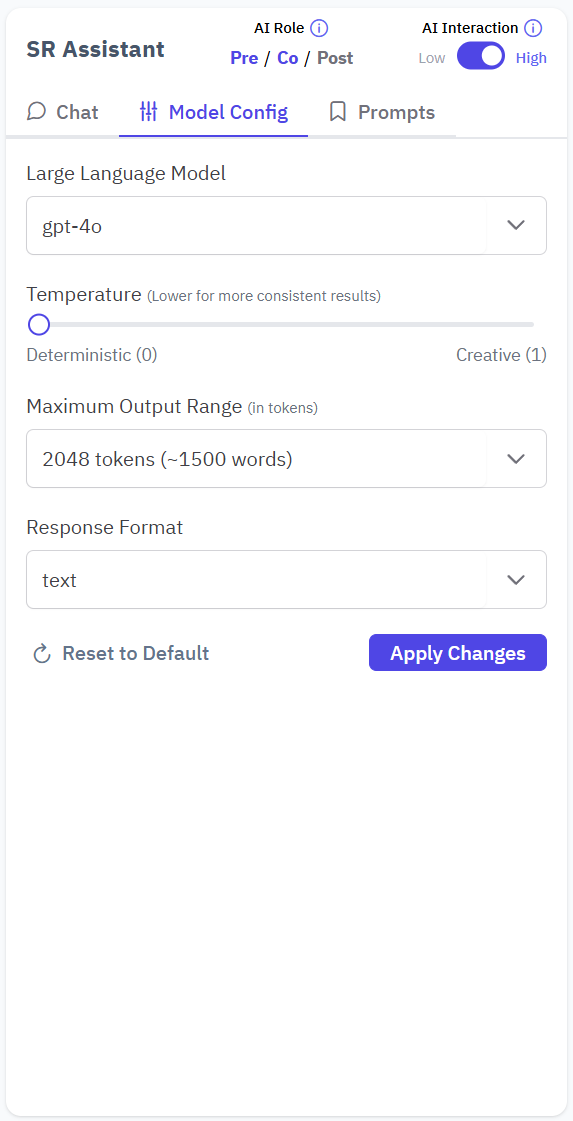}
		\caption{Model Config}
		\label{fig:config}
	\end{subfigure}
	\hspace{0.5pt}
	\begin{subfigure}[b]{0.1715\textwidth}  
		\includegraphics[width=\textwidth]{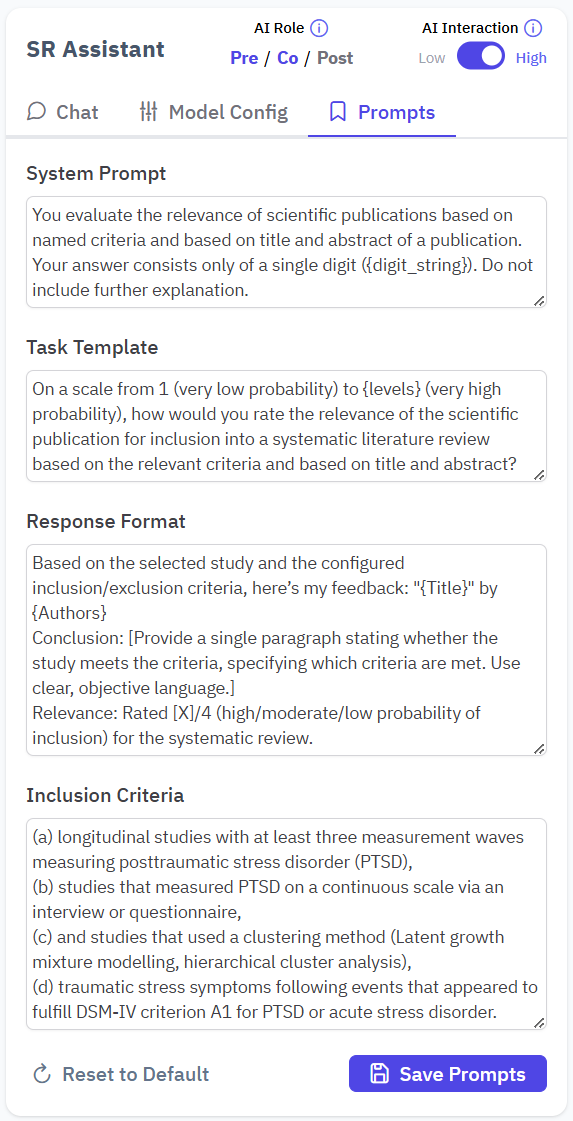}
		\caption{Prompts}
		\label{fig:prompts}
	\end{subfigure}
	\caption{In the screening interface (a), users can select studies to expand for abstract screening, indicated with a purple edge. They have the option to include or exclude the selected study~\circled{1}{Navy} for further detailed review. LLM suggestions~\circled{2}{Navy} are immediately visible, as the settings are configured for Pre-reviewer~\circled{5}{Navy} and high LLM interaction. When Co-reviewer~\circled{5}{Navy} is enabled, users can engage the SR assistant by clicking the `Ask AI' button \circled{3}{Navy}, which reveals interactive features within the right panel~\circled{4}{Navy}. This panel includes three tabs: `Chat', `Model Config' (b) and `Prompts' (c) for interacting with the LLM, adjusting model settings, and editing prompts, respectively. Users can start a new chat via~\circled{7}{Navy}. The response area~\circled{8}{Navy} has LLM feedback for inclusion based on the interaction level. In `low' mode, users are limited to interact with the LLM by predefined options, i.e., PICO Extraction and Detailed Reasoning~\circled{9}{Navy}, while in `high' mode~\circled{6}{Navy}, users can directly prompt the LLM using~\circled{10}{Navy}. The `Model Config' tab allows users to change the LLM model, temperature, and response settings. The `Prompts' tab enables users to edit LLM prompts about the objective and persona, instructions in a task template, response format, and inclusion criteria.}
\end{figure*}

\section{Introduction and Related Work}
Systematic reviews (SRs) are comprehensive literature reviews that identify and appraise relevant studies to answer targeted research questions. The most labour-intensive part of conducting a SR is title and abstract screening, where tens of thousands of studies (e.g., retrieved from Boolean search engines like PubMed) need to be screened, or in other words, assessed by humans~\cite{polanin2019best}. Screeners, typically medical researchers and librarians, judge each study based on predefined inclusion/exclusion criteria, often specified in terms of population, intervention, comparison, and outcome (PICO). Studies assessed as relevant are then re-assessed at the full-text level, so faster title and abstract screening can reduce the overall time of systematic review creation by running these two stages asynchronously. To accelerate title and abstract screening, open source tools such as ASReview~\cite{van2021open}\footnote{\url{https://asreview.nl}} and DenseReviewer~\cite{mao2025densereviewer}\footnote{\url{https://densereviewer.ielab.io/}} have emerged; especially with recent advances in Large Language Models (LLMs), alongside the emergence of commercial tools such as Elicit.%
\footnote{\url{https://elicit.com/solutions/systematic-reviews}. Note that Elicit does not perform title and abstract screening; instead, it uses LLMs for information extraction, supporting other downstream tasks.}
However, open source solutions--whether from developers or researchers--are usually distributed simply as raw scripts, posing a high entry barrier for screeners who conduct SRs but may lack expertise in the underlying techniques of these solutions. Commercial tools are often cautious in allowing users access to the underlying LLMs, and their implementation details are typically opaque, e.g., the prompts are not editable by end users. For example, among popular screening tools, only EPPI Reviewer currently specifies support for GPT-4-based data extraction and judgement, but it restricts this usage to post-screening evaluation scenarios.%
\footnote{\url{https://eppi.ioe.ac.uk/cms/Default.aspx?tabid=3921}}
Elicit claims to support SRs through text summarisation and intervention extraction but shows only the first few top-ranked studies.

We propose an analysis framework for categorising LLM usages by roles and interaction levels with human screeners to address the gap in the fact that no current SR software systematically integrates LLM-assisted screening. We present AiReview, a platform designed to (i) enable end users (e.g., medical researchers, librarians) with access to LLMs for SR tasks, e.g., title and abstract screening; and (ii) investigate the impact of LLMs on SR tasks, especially title and abstract screening. Our platform allows transparent control of LLMs, aligning with the guidelines for AI usage in SRs~\cite{cacciamani2023prisma}.
	\begin{table*}[t!]
	\caption{Roles and interaction patterns of LLMs in systematic review screening.}
	\centering
	\label{tab:llm-roles}
	\renewcommand{\arraystretch}{1.4}
	\begin{tabular*}{\linewidth}{
			>{\centering\arraybackslash}m{4.5cm}
			>{\centering\arraybackslash}m{4.5cm}
			>{\centering\arraybackslash}m{3.8cm}
			>{\centering\arraybackslash}m{3.8cm}
		}
		\toprule
		\multirow{4}{*}{\textbf{Role}} & 
		\multirow{4}{*}{\textbf{Workflow}} & 
		\multicolumn{2}{c}{\textbf{LLM Interaction Level}} \\
		& & \balanceindicator{1} & \balanceindicator{2} \\
		& & {\small Low support} & {\small High support} \\
		\midrule
		\textbf{Pre-reviewer} 
		\raggedright \newline Pre-screens studies with automated scoring and reasoning
		~\cite{dennstadt2024title,syriani2024screening,cao2024prompting,guo2024automated} & 
		{\faRobot}\,{\faLongArrowAltRight}\,{\doccheckx}\,\,\,{\faLongArrowAltRight}\,{\faUserMd} & 
		\parbox[b]{3cm}{\centering \faHandPointer[regular] \, Show results upon requested}  & 
		\parbox[b]{3cm}{\centering \faEye[regular] \, Reveal results along with studies}  \\
		\midrule
		\raggedright \textbf{Co-reviewer}
		\newline Provides live assistance during human screening
		~\cite{herbst2023accelerating,huotala2024promise,bron2024combining} & 
		{\faUserMd}\,{\faSync}\,{\faRobot}\,{\faLongArrowAltRight}\,{\doccheckx} & 
		\parbox[b]{2.5cm}{\centering \faHandPointer[regular] \, Help options for predefined tasks} & 
		\parbox[b]{3cm}{\centering \faComments[regular] \, In addition to options, enable chat} \\
		\midrule
		\raggedright \textbf{Post-reviewer}
		\newline Reviews human decisions and processes remaining studies
		~\cite{oami2024performance,huotala2024promise} & 
		{\faUserMd}\,{\faLongArrowAltRight}\,{\doccheckx}\,\,\,{\faLongArrowAltRight}\,{\faRobot} & 
		\parbox[b]{3cm}{\centering \faHandPointer[regular] \, Show LLM decisions for comparison} & 
		\parbox[b]{3cm}{\centering \faTasks  \, Check potential incorrect decisions, enable chat}  \\
		\bottomrule
	\end{tabular*}
\end{table*}

\section{Title and Abstract Screening with AiReview}
An overview of AiReview's architecture is shown in Figure~\ref{fig:arch}. At a high-level, users upload studies retrieved from PubMed in \texttt{nbib} format, along with the corresponding inclusion criteria for the SR as the input for LLM-assisted screening. After the screening, users can download the screened studies. Specifically, AiReview is deployed using two individual Docker containers, which can run on a single cloud instance. The front-end container delivers a web-based interface built with Vue.js and Tailwind CSS, with Nginx serving static content. The back-end container manages the system's core logic. It incorporates a PostgreSQL database for storing user-uploaded collections, screening results, and related data. RabbitMQ and Celery are employed for message queuing and asynchronous task management, enabling efficient handling of long-running processing requests and concurrent execution of multiple requests before http timeout. The core functionality of AiReview is implemented in Python, connecting to LLMs via external APIs to enable LLM-assisted screening. The front end communicates with the Python-based back end via REST APIs built with Django and WebSockets, enabling two-way communication between the user’s browser and the server for handling LLM streaming responses.

We present how users can leverage AiReview for their SR tasks. 
Figure~\ref{fig:main} illustrates the screening interface. The interface is divided into left and right panels.%
\footnote{The SR and studies for screening are from van de Schoot et al.~\cite{van2018bayesian}. GPT-4o is used as the LLM, with prompts adapted from Dennst{\"a}dt et al.~\cite{dennstadt2024title}.}
The left panel lists all uploaded studies, allowing users to decide inclusion or exclusion via green and red buttons. The right panel provides the SR Assistant, which acts as a copilot during screening. During the initial setup, users can choose the LLM roles they want to use and the desired level of LLM interaction for assisting the upcoming screening task. Here we present a case where screeners enable an LLM to suggest inclusion and exclusion decisions before screening (indicated as `Pre' in the AI Role) and activate an LLM to collaborate as a SR Assistant during screening (indicated as `Co' in the AI Role). The screening list of studies can be sorted directly by an LLM or from other recent approaches \cite{mao2024reproducibility,mao2024dense}. If LLM-assisted screening is enabled, users also need to provide inclusion/exclusion criteria. AiReview offers system prompts and task templates that users can edit and customise as needed.  During screening, users can interact with LLMs in the side panel, according to their predefined AI interaction preferences. After screening, users can review their decisions or compare them with LLM decisions (indicated as `Post' in the AI Role) and export the files (e.g., in \texttt{nbib} format) for downstream tasks.

Figure~\ref{fig:config} illustrates the model configuration interface. Users can customise the SR Assistant by selecting their preferred LLM and adjusting output style at the model level. AiReview supports both commercial 
(e.g. OpenAI GPTs\footnote{\url{https://platform.openai.com/docs/models}})
and open-source LLMs 
(e.g. Meta LLaMa series\footnote{\url{https://www.llama.com/docs/model-cards-and-prompt-formats/}}, 
Mistral AI\footnote{\url{https://docs.mistral.ai/getting-started/models/models_overview/}},
Deepseek\footnote{\url{https://api-docs.deepseek.com/quick_start/pricing}}), accessible via APIs. 
The SR Assistant is globally controlled by the AI Interaction switch at the system level, while response characteristics--such as diversity, length, and structure--can be configured here. 

Figure~\ref{fig:prompts} shows the prompt interface. AiReview loads predefined prompts for LLM-assisted screening, allowing users to check and edit them as needed. The listed prompts range from general to specific: System Prompt sets the basic instructions for the LLM, Task Template defines the task for LLM, Response Format controls the style of the response, and Inclusion Criteria is provided by users and controls the SR-Assistant screening.

	
\section{Framework for Categorising LLM Use Cases }
Table~\ref{tab:llm-roles} presents a framework we developed and applied for AiReview that systematically categorises LLM use cases in the screening task by role, workflow, and interaction level. LLMs serve three roles in the screening workflow: pre-reviewer, co-reviewer, and post-reviewer; or \textit{before}-, \textit{with}-, and \textit{after} human. We found that the LLM's place in the workflow affects the level of bias introduced. Conceptually, having LLMs make initial judgements will substantially impact human decisions, whereas positioning LLMs after human screening will see less influence~\cite{choi2024llm}. Finally, we discuss cases where screeners need different levels of support from the LLM to manage potential bias, which also guide our system design.

From this point, we introduce a `level of interaction' dimension (also referred to as collaboration integration~\cite{faggioli2023perspectives}), which subdivides each role based on the amount of support provided by the LLM (either high or low). Generally, the distinguishing factor between low and high interaction levels is whether the LLM's response is displayed to screeners (e.g., visibly shown \faEye[regular]) or triggered by screeners (e.g., a click \faHandPointer[regular]). Specifically, when the LLM is used as a pre-reviewer, recommendations and suggestions are already prepared before human screening. Thus, screeners desiring low LLM support must click the Ask AI button to display results for each study. If high support is desired, the results will be shown immediately upon entering the screening UI. When the LLM is used as a co-reviewer, screeners desiring lower LLM support have limited access to pre-defined LLM prompts. If high support is desired, screeners can freely chat with the LLM. When the LLM is used as a post-reviewer, the LLM can serve as a second reviewer to provide independent feedback. Similar to the pre-reviewer role, if low support is desired, the LLM will only show results for comparison. If high support is desired, the LLM will actively display comments based on screeners' decisions, and screeners can also freely chat with the LLM. 

\begin{table}[!t]
	\centering
	\caption{LLM Pipeline Categorisation with Effort Saved. Effort savings are conceptually deducted and represented using \faBolt{} symbols, where the Full Pipeline, which provides the highest level of automation and assistance, is marked with 7 \faBolt{}s. Other pipelines are assigned proportionally fewer \faBolt{}s based on their relative effort savings.}
	\label{tab:llm-pipelines}
	\resizebox{\columnwidth}{!}{ 
		\renewcommand\arraystretch{1.5}
		\begin{tabular}{llcccl}
			\toprule
			\textbf{Category} & \textbf{Pipeline} & \textbf{Pre} & \textbf{Co} & \textbf{Post} & \textbf{Effort Saved} \\
			\midrule
			Decision-making & Pre-Only & \centering \tiny\faCheck  & \centering \tiny\faTimes  & \centering \tiny\faTimes  & 
			\faBolt{} \faBolt{} \faBolt{} \\
			\cmidrule{1-6}
			\multirow{2}{*}{Live Collaboration} 
			& Pre-Co Pipeline & \centering \tiny\faCheck  & \centering \tiny\faCheck  & \centering \tiny\faTimes  & 
			\faBolt{} \faBolt{} \faBolt{} \faBolt{} \faBolt{} \faBolt{} \\
			\cline{2-6}
			& Co-Only & \centering \tiny\faTimes  & \centering \tiny\faCheck  & \centering \tiny\faTimes  & \faBolt{} \faBolt{}\\
			\cmidrule{1-6}
			\multirow{3}{*}{Quality Control} 
			& Pre-Post Pipeline & \centering \tiny\faCheck  & \centering \tiny\faTimes  & \centering \tiny\faCheck  & 
			\faBolt{} \faBolt{} \faBolt{} \faBolt{} \faBolt{} \\
			\cline{2-6}
			& Co-Post Pipeline & \centering \tiny\faTimes  & \centering \tiny\faCheck  & \centering \tiny\faCheck  & 
			\faBolt{} \faBolt{} \faBolt{} \faBolt{} \\
			\cline{2-6}
			& Post-Only & \centering \tiny\faTimes  & \centering \tiny\faTimes  & \centering \tiny\faCheck  & 
			\faBolt{} \\
			\cmidrule{1-6}
			Full Assistance & Full Pipeline & \centering \tiny\faCheck  & \centering \tiny\faCheck  & \centering \tiny\faCheck  & 
			\faBolt{} \faBolt{} \faBolt{} \faBolt{} \faBolt{} \faBolt{} \faBolt{} \\
			\bottomrule
	\end{tabular}}
\end{table}

\vfill\eject
AiReview supports both single and composable pipelines through the use of the three LLM roles identified above: `pre-reviewer' (P), `co-reviewer' (C), and `post-reviewer' (Q). We categorise the valid use cases---seven in total (P; C; Q; (P, C); (P, Q); (C, Q); (P, C, Q))---along with their associated conceptual effort savings in Table~\ref{tab:llm-pipelines}. Note that screeners are only able to use the post-reviewer role once all studies have been screened, and the LLM does not use any of the information from the screener to make assessments. Thus, the pre-reviewer role is designed to assist with decision-making, while the post-reviewer is designed to assist with quality control.

We estimate the effort saved by each LLM pipeline by reasoning about the amount of manual effort reduced by each role. First, we hypothesise that applying all three roles in conjunction with one another---pre-reviewer (\(P\)), co-reviewer (\(C\)), and post-reviewer (\(Q\))---saves more effort than using any other combination of roles: \((P, C, Q) \succ P \lor C \lor Q\) and \((P, C, Q) \succ (P, C) \lor (C,Q) \lor (P, Q)\).  Pre-reviewer (\(P\)) reduces workload before human involvement, whereas co-reviewer (\(C\)) only assists without actively replacing manual effort, leading to \(P \succ C\). In contrast, post-reviewing (\(Q\)) merely validates human decisions and does not impact initial manual effort, meaning it inherently assumes prior human involvement. Consequently, post-reviewer is the least effective in reducing conceptual effort, leading to \(P \succ Q\) and \(C \succ Q\). Between the two combinations, \((P, C)\) reduces effort both before and during screening, whereas \((C, Q)\) only assists and validates post-screening, establishing \((P, C) \succ (C, Q)\). Additionally, \((P, Q)\) saves effort before screening like  \((P, C)\), but lacks live assistance, meaning \((P, C) \succ (P, Q)\). Since suggesting assessments before screening is more effective than during screening,we also establish \((P, Q) \succ (C, Q)\). Summarising these relationships:
\[
(P, C, Q) \succ (P, C) \succ (P,Q) \succ (C, Q) \succ P \succ C \succ Q.
\]
We present the ranking with \faBolt[regular]s in Table~\ref{tab:llm-pipelines}. We also hypothesise that the effort savings correlate with the bias introduced by LLMs: the more human effort saved, the more bias is introduced.

\vfill\eject
Based on the pipelines in Table~\ref{tab:llm-pipelines}, Table~\ref{tab:llm-cases} illustrates three scenarios where AiReview addresses specific needs in real-world settings. 
When students use AiReview to learn how to do T\&A screening, the teaching team can activate the co-reviewer mode, allowing students to seek help from the SR Assistant. Students can critically evaluate the AI's suggestions and learn from the process, even when AI hallucinates and provides incorrect information. Similarly, AiReview supports screening teams with varying sizes: as a second screener in resource-limited settings for timely progress or as an additional screener to ensure quality control.

This analysis framework is not only applicable to title and abstract screening, but can also generalise to other tasks where LLMs have potential to help. For example, in query formulation, LLMs can serve as a pre-, co-, or post-builder for Boolean queries~\cite{wang2023can}, the main way studies are retrieved for systematic reviews.

	\begin{table}[t]
	\centering
	\caption{Real-World Use Cases of LLM-Assisted Systematic Review Screening}
	\label{tab:llm-cases}
	\resizebox{\columnwidth}{!}{
		\begin{tabular}{p{3.5cm}p{1.5cm}p{4.6cm}}
			\toprule
			\textbf{Scenario} & \textbf{Suggested Pipeline(s)} & \textbf{Illustrative Example} \\
			\midrule
			\raggedright
			\faGraduationCap{} \textbf{Students Learning to Screen}  \linebreak
			New researchers receive real-time feedback while screening.  
			
			&  
			\textbf{Co-Only Pipeline}  
			
			&  
			\faCommentDots{} \textbf{User}:
			"I am unsure if this study meets the PICO criteria. Can you provide feedback?"
			
			\faRobot{} \textbf{LLM}:
			"The study mentions the correct population but lacks details on the intervention. Check the Methods section."
			\\  
			\midrule
			
%
%
%
			\raggedright
			\faBalanceScale{} \textbf{Resource-Limited Teams}\linebreak
			Teams with fewer screeners use the LLM as an additional reviewer for consistency.  
			
			&  
			\textbf{Full Pipeline}  
			
			&  
			\faCommentDots{} \textbf{User}:  
			"We have only one reviewer. Can you act as a second screener and provide justifications?"
			
			\faRobot{} \textbf{LLM}:  
			"For this paper, I recommend inclusion based on the intervention. The next paper lacks a comparator and should be excluded."
			\\  
			\midrule
			\raggedright
			\faCheckCircle{} \textbf{Quality Control\linebreak After Screening} \linebreak 
			LLM identifies inconsistencies post-screening, ensuring criteria adherence.  
			
			&  
			\textbf{Co-Post Pipeline}  
			
			&  
			\faCommentDots{} \textbf{User}:  
			"Can you review included studies and highlight any inconsistencies?"
			
			\faRobot{} \textbf{LLM}:  
			"I noticed that two similar studies were handled differently. Do you want to revisit this decision?"
			\\  
			\bottomrule
	\end{tabular}}
\end{table}
\section{Conclusion and Future Work}
In this paper, we introduce AiReview and our analysis framework for LLM use cases, showcasing T\&A screening as an example of a LLM-assisted tool. We plan to conduct a user study using this platform to investigate how different roles and interaction levels of LLMs affect human's screening decision and perceived utility, and if it can benefit screeners with LLM usage budgets. In the future, we will expand this platform by including LLM use cases for other SR tasks, such as Boolean query formulation and data extraction, to explore the possibility of building an end-to-end solution for systematic review creation. Additionally, we aim to scale beyond current individual screening to support collaborative screening with multiple team members, leveraging LLMs to assist in resolving disagreements. Finally, we will investigate whether LLM-assisted screening can benefit from previous non-LLM ranking methods, such as DenseReviewer~\cite{mao2025densereviewer}.


\bibliographystyle{ACM-Reference-Format}
\bibliography{Bib/bibliography}
\end{document}